\documentclass[pre,amsmath,amssymb]{revtex4}

\usepackage{graphicx}
\usepackage{dcolumn}
\usepackage{bm}
\usepackage[latin1]{inputenc}   
\usepackage[T1]{fontenc}        

\begin{document}


\title{Breakdown of disordered media by surface loads}

\author{Jakob Knudsen}
\affiliation{%
  Materials Science, Malmö University, SE 205 06 Malmö, Sweden
}%
\affiliation{ Solid Mechanics, Luleå University of Technology, SE 971
  87 Luleå, Sweden
}%

\author{A. R. Massih}
\affiliation{%
  Materials Science, Malmö University, SE 205 06 Malmö, Sweden
}%
\affiliation{ Quantum Technologies, Uppsala Science Park, SE 751 83
  Uppsala, Sweden
}%

\date{\today}

\begin{abstract}
  We model an interface layer connecting two parts of a solid body by
  $N$ parallel elastic springs connecting two rigid blocks.  We load
  the system by a shear force acting on the top side. The springs have
  equal stiffness but are ruptured randomly when the load reaches a
  critical value.  For the considered system, we calculate the shear
  modulus, $G$, as a function of the order parameter, $\phi$,
  describing the state of damage, and also the ``spalled'' material
  (burst) size distribution. In particular, we evaluate the relation
  between the damage parameter and the applied force and explore the
  behaviour in the vicinity of material breakdown.  Using this simple
  model for material breakdown, we show that damage, caused by applied
  shear forces, is analogous to a first-order phase transition.  The
  scaling behaviour of $G$ with $\phi$ is explored analytically and
  numerically, close to $\phi=0$ and $\phi=1$ and in the vicinity of
  $\phi_c$, when the shear load is close but below the threshold force
  that causes material breakdown. Our model calculation represents a
  first approximation of a system subject to wear induced loads.
\end{abstract}

\maketitle


\section{Introduction}
\label{sec:intro}

The damage of solids subject to external loads is a progressive
physical process, which eventually may lead to material breakdown. At
the microscopic scale, damage is caused by the localisation of the
stress field in the neighbourhood of defects or at interfaces that
induce the breaking of the atomic bonds. At a mesoscopic scale there
is the interaction and coalescence of microscopic cracks or vacancies
(voids), which together initiate a macroscopic crack.

One category of material damage problems, of great technological and
fundamental interest, is the deterioration of solid surfaces in
contact induced by relative motion. These problems belong to the field
of tribology, in which friction and wear of materials are studied
\cite{Bowden_Tabor_2000,Rabinowicz_1995}. The first phenomenological
approach to the problem of wear dates back to the 1940's, when Holm
\cite{Holm_1946} provided a simple description for wear of solids,
which later was re-derived and applied by Archard \cite{Archard_1953}.
The Holm-Archard wear description states that the amount of wear or
the worn mass between two rubbing surfaces is proportional to sliding
work (or dissipation energy due to friction) divided by the hardness
of the surface; simply $m \propto F_Nx/p$, where $m$ is the worn mass,
$F_N$ the normal force or the force applied normal to the interface,
$x$ the distance slid and $p$ is the penetration hardness, which is
related to the surface tension of the solid, $\gamma$, in the manner
$p \propto \gamma^3$, see \cite{Rabinowicz_1995}. A mass loss from
a solid body associates with heat dissipated to the environment, or
from thermodynamics, $dQ \le TdS$, where $dQ$ is the increment of heat
exchange, $T$ the temperature and $dS$ is the change in entropy. A
shear force, $F_S$, applied to one of the bodies in a direction
parallel to the interface plane, causes slip (a relative motion
between parts of the interface) when $F_S=\mu F_N$, where $\mu$ is the
coefficient of friction.  This is Amontons's second law of macroscopic
friction, formulated more than 200 years ago \cite{Bowden_Tabor_2000}.
Setting $dQ=\mu F_N dx$ and $\mu=G/p$, with $G$ denoting the shear
strength of softer material, we can express the Holm-Archard
description as $dm \le TdS/G$, \emph{i.e.} the worn mass is
proportional to the production of entropy and inversely to the shear
strength of softer material.

The aim of our study is to investigate the manner in which a solid
material breaks down, when it is subjected to a shear force that
reaches a certain critical value, by using a simple model for material
damage.  The system under study is described by $N$ parallel elastic
springs connecting two rigid blocks.  The system is loaded with a
shear force acting on the top side. If the relative motion is slow
and the system is well cooled, it is possible to neglect the heat
generated at the contact. Furthermore, neglecting the influence of the
normal load, which mainly introduces compressive stresses, and
supposing that the crack path is limited to the interface between the
solid blocks, we arrive at the system studied in this paper. The
model can represent rupture of a weld between two solid bodies due to
shear forces.

A solid body subject to stress can be considered as being in a
metastable state. It can transform in a self-organised manner to
stable fracture state by formation of cracks. This self-organised
behaviour of de-bonding in solids, which is the start of the damage
process, is theoretically analogous the phenomenon of nucleation in
first-order phase transition
\cite{Selinger_et_al_1991,Buchel_Sethna_1996,Buchel_Sethna_1997}.
Zapperi \emph{et al.} \cite{Zapperi_et_al_1997,Zapperi_et_al_1999}
have utilised this analogy to study the global breakdown of disordered
media under external loads. They used a two-dimensional lattice model,
in which each bond of the lattice represents an elastic spring that
breaks when it is stretched beyond a threshold value governed by a
probability distribution. They studied numerically the random fuse
model (RFM) and a spring network in the framework of a mean-field
theory. They observed that the breakdown is preceded by avalanche
events, which is analogous to the formation of the droplets observed
near a spinodal decomposition
\cite{Klein_Unger_1983,Unger_Klein_1985}.

Lattice or discrete modelling approach to material damage dates back
to the 1926 work of Peirce, what is presently known as the global
sharing fibre bundle \cite{Peirce_1926}. Peirce considered $N$
parallel fibres, each with its own rupture threshold. The fibres were
linked in such a way that when one link failed, the load was
transformed to and shared equally among all the intact links. Later,
Daniels \cite{Daniels_1945} studied and extended this model in detail
and determined the asymptotic distribution of the bundle strength for
large $N$. Sornette \cite{Sornette_1989} evaluated the failure
characteristic of $N$ independent vertical lines linked in parallel
with identical spring constants and random failure thresholds. He
showed that the rupture properties of this system is quite distinct
from systems that are linked in series. Furthermore, Sornette
identified different regimes of rupture depending on the range of
stresses. (i) An elastic-reversible behaviour was observed for
stresses below certain threshold, $\sigma_1$ at which the first link
failed, (ii) a stress range $\sigma_1 \le \sigma_N$, where the system
exhibited a non-linear elastic but reversible behaviour, and finally,
(iii) for $\sigma \ge \sigma_N$ global failure occurred. Moreover,
Sornette showed that the asymptotic properties of global failure
threshold of the network system can be explained in terms of the
central limit theorem of statistics. A review of rupture models can be
found in Sornette's book \cite{Sornette_2000}. The book by Hermann and
Roux offers a review of the field up to 1990 \cite{Hermann_Roux_1990}.
Hemmer and Hansen \cite{Hemmer_Hansen_1992} studied and calculated the
distribution of burst avalanches in fibre bundles. They found that for
a large class of failure threshold distributions, the bursts are
distributed according to an asymptotic scaling law $s^{-5/3}$ for
simultaneous rupture of $s$ fibres.

In this paper, we pursue a similar approach as Zapperi \emph{et al.}
\cite{Zapperi_et_al_1999} to investigate analytically and numerically
the breakdown of disordered media subject to surface loads. We
envisage that the solid, subjected to shear loading, contains an
interface layer, which consists of a network of parallel springs. We
study the behaviour of the shear modulus, the order parameter
characterising the state of damage and burst (avalanche) size
distribution.  We compare the universality of our results with some
other failure models, \emph{e.g.} the fibre bundle models and discuss
our results in the context of the magnetic system.

The organisation of this paper is as follows: In section
\ref{sec:model}, we present the physical model and its associating
constitutive relations. In section \ref{sec:method}, we outline the
method of computation, numerical and analytical, for the evaluation of
the shear modulus, the variation of the applied load with the order
parameter, the critical exponents and the avalanche size distribution.
The results of our computations are presented in section
\ref{sec:result} and discussed in section \ref{sec:discus}. Some
relating background material is placed in the appendix.


\section{Model}
\label{sec:model}

\begin{figure}[htbp]
  \centering \includegraphics[]{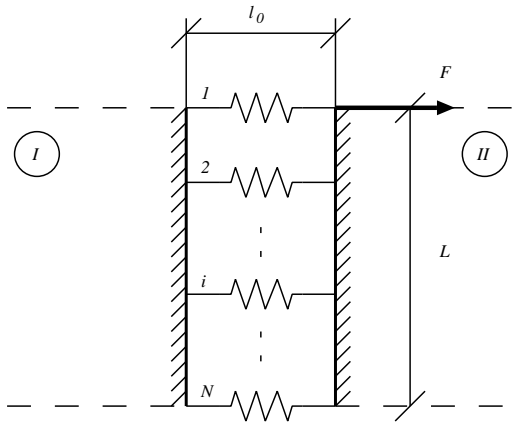}
  \caption{$N$ parallel springs, connecting bodies I and II, subjected to
    a shear load $F$.}
  \label{fig:parallell_model}
\end{figure}

Consider an interface layer connecting two parts of a solid body
consisting of $N$ equi-spaced parallel springs of equal length, $l$,
with spring constants, $k_j$, and\emph{ randomly distributed} rupture
deformation thresholds, $D_j$, where $j \in [1,N]$. Initially the
springs constants will be set equal ($k_j = k$). If a spring is
stretched beyond its rupture threshold, it will loose its load
carrying capacity, \emph{i.e.}, $k_j = 0$.  The springs are loaded
with a force, $F$, acting parallel to the upper surface simulating the
shear component of a wear-inducing load, Fig.
\ref{fig:parallell_model}. Each spring carries the load, $F_j = k_j
\Delta_j$, where $\Delta_j = l_j - l$ denotes the displacement of
spring $j$.  Assuming linear deformation, the deformation of spring
$i$ can be written as

\begin{equation}
  \label{eq:spring_displacement}
  \Delta_i = \frac{N-i}{N-1}\Delta_1 + \frac{i-1}{N-1}\Delta_N,
\end{equation}
\noindent
where $\Delta_1$ and $\Delta_N$ is the displacement of the end
springs. The state of equilibrium gives

\begin{equation}
  \label{eq:equilib_state}
  F=\sum_{i=1}^N F_i=\sum_{i=1}^N k_i \Delta_i;\qquad  \sum_{i=2}^N
  F_i(i-1)a=0,
\end{equation}
\noindent
where the equation on the right states that the sum of the moments of
all the forces acting on the body is zero and $a$ denotes the lattice
spacing. Equations~(\ref{eq:spring_displacement}) and
(\ref{eq:equilib_state}) are combined to find the boundary
displacements

\begin{eqnarray}
  \Delta_1 & = & F(N-1) \left\{ 
    \sum_{i=1}^N k_i (N-i) - 
    \frac{ \sum_{i=2}^N k_i (i-1)(N-i) }{ \sum_{i=2}^N k_i (i-1)^2 }
    \sum_{i=1}^N k_i (i-1) 
  \right\}^{-1}, \label{eq:displacement_force_1} \\
  \Delta_N & = & - \Delta_1 
  \frac{ \sum_{i=2}^N k_i (i-1)(N-i) }{ \sum_{i=2}^N k_i (i-1)^2 }.
  \label{eq:displacement_force_N}
\end{eqnarray}

Let us introduce the dimensionless variables, $\kappa_i = k_i/k$,
$\varepsilon_i = \Delta_i/l$, $r_i = D_i/l$ and $f = F/( k l N )$,
then Eq.~(\ref{eq:spring_displacement}) gives the expression for the
strain of spring $i$, \textit{ viz.}

\begin{equation}
  \label{eq:strain_force_1}
  \varepsilon_i = N f K_i \equiv Nf \left[ \frac{X_i^N}{\sum_{j=1}^N
      \kappa_jX_j^N}\right], 
  \end{equation}
\noindent
where
\begin{equation}
  \label{eq:strain_force_2}
 X_i^N = (N-i)-S_N(i-1);\qquad S_N  \equiv \frac{ \sum_{i=2}^N
 \kappa_i (i-1)(N-i) }{ \sum_{i=2}^N \kappa_i (i-1)^2 },
\end{equation}
\noindent
with properties $X_N^N=-S_NX_1^N$ and
$\kappa_i=\Theta(1-\varepsilon_i/r_i)$, where $\Theta(x)=1$ for $x>0$
and $\Theta(x)=0$ for $x<0$.
    
We now express the total dissipated energy due to breaking of springs

\begin{equation}
  \label{eq:energy_dissipation_dim}
  E = \frac{1}{2} \sum_{i=1}^N k_i \left( \Delta_i^2 - D_i^2 \right),
\end{equation}
\noindent
which can be rewritten in terms of the dimensionless variables in the
form
\begin{equation}
  \label{eq:energy_dissipation}
  \mathcal{E}\{\kappa\} \equiv \frac{E}{k l^2} = \frac{1}{2} \sum_{i=1}^N \kappa_i \left(
    \varepsilon_i^2 -r_i^2 \right).
\end{equation}

We further rewrite this equation in terms of the external applied
force $f$ using Eq.~(\ref{eq:strain_force_1}), \textit{viz.}

\begin{equation}
  \label{eq:energy_force}
  \mathcal{E}\{\kappa\} = \frac{1}{2} \left[  \frac{N f^2}{G\{\kappa\}} - 
  \sum_{i=1}^N \kappa_i r_i^2 \right],
\end{equation}
\noindent
where $G\{\kappa\}$ is given by

\begin{equation}
  \label{eq:shear_modulus}
  G\{\kappa\} = \left[N\sum_{j=1}^N  \kappa_j K_j^2\right]^{-1}.
\end{equation}

Here, $G$ can be interpreted as a global shear modulus of the lattice.
It depends on the number of broken springs and the order in which the
springs break.

We define the order parameter for the system
\begin{equation}
  \label{eq:damage_order}
  \phi = \frac{1}{N}\sum_{i}^N \kappa_i.
\end{equation}

Then Eq. (\ref{eq:energy_force}) is expressed in the form

\begin{equation}
  \label{eq:mean_field_2}
  \mathcal{E} =  \frac{1}{2} \sum_{i=1}^N \kappa_i 
  \left( \frac{f^2}{\phi G(\phi)} - r_i ^2 \right), 
\end{equation}
\noindent
where $G(\phi)$ is re-expressed as a function of the order parameter.
As has been shown by Zapperi \emph{et al.} \cite{Zapperi_et_al_1999},
the self-consistency requirement on $\phi$ yields the following
integral equation (see Appendix)
\begin{equation}
  \label{eq:order_param_4}
  \phi = 1-\int_{0}^{\frac{f}{\sqrt{\phi G(\phi)}}}
  \psi(r)\mathrm{d}r,
\end{equation}
\noindent
where $\psi(r)$ is the probability distribution function for the
system breakdown.

The susceptibility for the system breakdown is then calculated
according to
\begin{equation}
  \label{eq:chi}
  \chi \equiv \frac{d\phi}{df} = -\frac{y\psi(y)}{f-y^2\psi(y)h_\phi},
\end{equation}
\noindent
where
\begin{equation}
  \label{eq:par}
  y = \frac{f}{\sqrt{\phi G(\phi)}}; \qquad
  h_\phi = \frac{1}{2}\frac{G(\phi)+\phi G'(\phi)}{\sqrt{\phi G(\phi)}},
\end{equation}
\noindent
with $G'=dG/d\phi$. At the onset of breakdown, $f=f_c$ and
$\phi=\phi_c$, the susceptibility diverges; this corresponds to
\begin{equation}
  \label{eq:chi_critical}
 y_c \psi(y_c) = \frac{2\phi_c G(\phi_c)}{G(\phi_c)+\phi_c G'(\phi_c)}, 
\end{equation}
with $y_c=f_c/(\phi_c G(\phi_c))^{1/2}$.  We note that by specifying
the form of the distribution function $\psi$,
Eq.~(\ref{eq:chi_critical}) determines $G(\phi_c)$. In particular,
for $\psi = c$, where $c$ is a constant, the differential
equation~(\ref{eq:chi_critical}) can be solved to give
\begin{equation}
  \label{eq:G_critical}
  G( \phi_c ) = \frac{ c^2 f_c^2 }{ \phi_c \left( \phi_0 - \phi_c \right)^2
  },
\end{equation}
\noindent
where $\phi_0>\phi_c$ is an integration constant, which can be
determined by computation.

For $f<f_c$, in the vicinity of $f_c$, the susceptibility scales as
\cite{Zapperi_et_al_1999},
\begin{equation}
  \label{eq:chi_scale}
  \chi= \frac{d\phi}{df}\sim (f_c-f)^{-1/2}.
\end{equation}

Similarly, the corresponding description for the order parameter is
\begin{equation}
  \label{eq:par_scale}
  \phi-\phi_c \sim (f_c-f)^{1/2}.
\end{equation}

Both relations (\ref{eq:chi_scale}) and (\ref{eq:par_scale}) are the
mean-field theory predictions, see Appendix.

The shear modulus $G(\phi)$ is expected to obey, in general, a
universal scaling law for $\phi\approx 0$, in the form

\begin{equation}
  \label{eq:G_0}
  G(\phi)=G_0\Theta(\phi-\frac{1}{N})\phi^\tau,
\end{equation}
\noindent
where $G_0$ is a material constant and $\tau$ is a universal constant
determined by experiment or computation. It is determined by
re-arranging Eq.~(\ref{eq:G_0})
\begin{equation}
  \label{eq:critical_exponent}
  \tau = \lim_{ \phi \rightarrow 0}
    \frac{\log|G(\phi)|}{\log|\phi|}.  
\end{equation}

Finally in our analysis, we consider the concept of burst avalanche
distribution \cite{Hemmer_Hansen_1992}. This concept has relevance to
a system undergoing wear, for which the size of the avalanches can be
used to estimate the size of the wear fragments. The size of a burst
avalanche, $s$, in our system is defined as the number of springs that
break simultaneously for a constant load \cite{Batrouni_et_al_2001}.
The burst size distribution, $\mathcal{N}(s)$, is generated by
counting the number of bursts of size $s$ that occur as the system
breaks down.  Hemmer and Hansen \cite{Hemmer_Hansen_1992} showed that
for a large class of failure threshold distributions in the fibre
bundle models, the bursts are distributed according to an asymptotic
power law,

\begin{equation}
  \label{eq:burst_critical}
   \lim_{ N \rightarrow \infty}
    \frac{\mathcal{N}(s)}{N}\sim s^{-\xi},
\end{equation}
\noindent
with a universal exponent, $\xi=5/2$.

\section{Computation method}
\label{sec:method}

The mean-field theory presented above requires as input the shear
modulus, $G$.  Once the state of the system is known, \emph{i.e.} the
number and position of the ruptured links is known, the shear modulus
of the system can be determined from Eq.~(\ref{eq:shear_modulus}).
However, the position of the ruptured springs depends on the sequence
in which they rupture. The sequence, in turn, depends on the local
rupture thresholds and the applied force. Even for moderate system
sizes ($N \sim 1000$) the number of possible sequences to failure
become prohibitively large, hence a statistical method is needed to
generate an average, or ``most probable'', shear modulus curve as a
function of the order parameter $\phi$. We take two different
approaches to accomplish this; (\emph{a}) a numerical scheme and
(\emph{b}) an analytic procedure. We compare the mean-field
predictions with quasi-static simulations of the lattice breakdown.

\subsection{Numerical}
\label{sec:numerical_method}

The probability that the spring $i$ breaks first under a uniform
strain field is $p = 1/N$, since the rupture limits are selected from
the same distribution. Here the springs are subjected to a linear
strain field, meaning that the probability that the spring $i$ breaks
first, can be written in the form

\begin{equation}
  \label{eq:first_spring_prob}
  p_i = w_i p, \quad \mathrm{with} \quad w_i = 
  \frac{ \epsilon_i^{+} }{ \sum_{i=1}^{N} \epsilon_i^{+} },  
\end{equation}

\noindent
where $w_i$ is a weighting factor based on the local strains
$\epsilon_i^+$ of the springs and the plus superscript indicates that
only positive strains are counted, \emph{cf.}
Eq.~(\ref{eq:strain_force_1}).

\begin{figure}[htbp]
  \centering \includegraphics[]{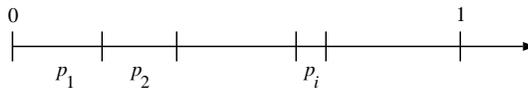}
  \caption{Illustration of how the sequence of breaking of springs is
    determined from  a random number generated according to a uniform
    distribution, see Eq.~(\ref{eq:first_spring_prob}).}
  \label{fig:spring_order}
\end{figure}

The sequence of spring ruptures is then determined numerically by
selecting a random number from a uniform distribution in the interval
$[0,1]$, and breaking the corresponding spring, see
Fig.~\ref{fig:spring_order}. The rupture probabilities, according to
Eq.~(\ref{eq:first_spring_prob}), are then re-computed and a new
rupturing spring is selected. The procedure is repeated until a
complete path to total failure is established.  The average shear
modulus curve is then found by averaging over several such sequences.
Finally, simple regression analysis is made to find a suitable
polynomial or power law expression approximating the average shear
modulus curve.

A\emph{ quasi-static} method is used to perform the numerical
simulations of the breakdown of the spring system
(Fig.~\ref{fig:parallell_model}).  The simulation is initiated by
associating a rupture limit to each individual spring in the system.
The limits are chosen from a predefined distribution. A full set of
rupture limits is called a configuration. For each configuration, the
computation starts with all links intact, $\kappa_i = 1$, and the
external load equal to zero, $f = 0$. Next, the load is incremented
and the local strains are computed according to
Eq.~(\ref{eq:strain_force_1}). The local strains are then compared to
their respective rupture limits. If the local strain exceeds the
rupture limit, then the link is broken, \emph{i.e.} the local
stiffness is set equal to zero. If a rupture has occurred, the local
strains are re-computed and checked once more against the rupture
thresholds, otherwise the computation proceeds with the next load
increment. When only one spring remains, the load loop is interrupted
and the computation proceeds to the next configuration.

Three different failure distribution functions are studied here;
(\emph{i}) a uniform distribution in the interval $r_i \in [ 0; 2 ]$,

\begin{equation}
  \label{eq:uniform}
  \psi(r) = \left\{ 
    \begin{array}[c]{ll}
      \frac{1}{2} & \quad r \in [0,2] \\
      0 & \quad \mathrm{otherwise}
    \end{array}
    \right.
\end{equation}

\noindent (\emph{ii}) a Weibull distribution with shape factor $\nu
= 2$ and location parameter $\lambda = 1$,

\begin{equation}
  \label{eq:weibull}
  \psi(r) = \left\{ 
    \begin{array}[c]{ll}
      \frac{\nu}{\lambda} \left( \frac{r}{\lambda} \right)^{\nu-1} 
      e^{ (r/\lambda)^\nu } & \quad r \ge 0 \\
      0 & \quad r < 0
    \end{array}
  \right.
\end{equation}

\noindent (\emph{iii}) a Gaussian (normal) distribution with mean value
$\vartheta=1$ and standard deviation $\sigma=$1,

\begin{equation}
  \label{eq:gauss}
  \psi(r) = \frac{1}{\sigma\sqrt{2\pi}} e^{-(r-\vartheta)^2/2\sigma^2}. 
\end{equation}

The critical values, $\phi_c$ and $f_c$, can be estimated directly
from the mean-field analysis from Eq.~(\ref{eq:chi_critical}). For the
uniform distribution this equation is easily solved, leading to a
functional relationship between $f_c$ and $\phi_c$. For the Weibull
and normal distributions, a Newton-Raphson scheme is employed to find
the load corresponding to a given value of the order parameter, by
finding the roots of Eq.~(\ref{eq:chi_critical}). The result is a line
in the ($\phi_c$, $f_c$) plane, which can be compared with the
critical values from the quasi-static simulations.

\subsection{Analytical}
\label{sec:method:analytic}

The material constant $G_0$, \emph{cf.}  Eq.~(\ref{eq:G_0}), is
determined exactly when all the springs are intact, \emph{i.e.} at
$\phi = 1$. For this case, Eq.~(\ref{eq:shear_modulus}) reduces to the
following simple relation:

\begin{equation}
  \label{eq:material_constant}
  \left.G\right|_{\phi=1} = G_0 = \frac{1+N}{-2+4 N}.
\end{equation}

Hence, as $N \rightarrow \infty$, $G_0 \rightarrow 1/4$.  The exact
expression for $G$ consists of terms like $\sum_{i=1}^{N}\kappa_i$,
$\sum_{i=1}^{N}\kappa_i i$ and $\sum_{i=1}^{N}\kappa_i i^2$. With only
one spring broken, denoted by $j$, the sums can be evaluated

\begin{eqnarray}
  \label{eq:one_broken}
  \sum_{i=1}^{N}\kappa_i & = & N - 1, \\
  \label{eq:two_broken}
  \sum_{i=1}^{N}\kappa_i i & = & N^2/2 + N/2 - j, \\
  \label{eq:three_broken}
  \sum_{i=1}^{N}\kappa_i i^2 & = & N^3/3 + N^2/2 + N/6 - j^2.
\end{eqnarray}

Hence, the slope of $G$ for $\phi\approx 1$ can be determined, once
the position of the first spring rupture is known, \emph{viz.}

\begin{equation}
  \label{eq:shear_slope}
  \left.\frac{\mathrm{d}G}{\mathrm{d}\phi}\right|_{\phi = 1} \approx
  \frac{G(1) - G(1-1/N)}{1/N}.
\end{equation}

We note that the first spring breaks according to a linear
distribution, \emph{cf.} Eq.~(\ref{eq:first_spring_prob}), which is
identified by $j = \mathrm{int}\{2(N-1)/9\}$. It corresponds to the
average value of the linear distribution. Inserting this $j$ into
Eqs.~(\ref{eq:two_broken})-(\ref{eq:three_broken}), then evaluating
Eq.~(\ref{eq:shear_slope}) and letting $N \rightarrow \infty$ leads to
$\left.\frac{\mathrm{d}G}{\mathrm{d}\phi}\right|_{\phi = 1} =
\frac{4}{9}$. Consequently, the following linear approximation of $G$
is found

\begin{equation}
  \label{eq:analytic_high}
  G_{\phi\approx 1} = \frac{4}{9}\phi - \frac{7}{36}.
\end{equation}

For the system under study, the $N$th spring is always the last to
break, since it is subject to compressive load. Moreover, the shear
modulus is zero when only one spring remains intact. When only two
intact springs remain in the system, denoting the next to last spring
to break as $m$, the sums in $G$, \emph{cf.}
Eqs.~(\ref{eq:one_broken})-(\ref{eq:three_broken}), are evaluated:

\begin{eqnarray}
  \label{eq:two_left}
  \sum_{i=1}^{N}\kappa_i & = & 2, \\
  \label{eq:three_left}
  \sum_{i=1}^{N}\kappa_i i & = & N + m, \\
  \label{eq:four_left}
  \sum_{i=1}^{N}\kappa_i i^2 & = & N^2 + m^2.
\end{eqnarray}

Inserting Eqs.~(\ref{eq:two_left})-(\ref{eq:four_left}) in Eq.
(\ref{eq:shear_modulus}) and using $\phi = 2/N$, it is possible to
carry out the limit analysis in Eq.~(\ref{eq:critical_exponent}), once
the position of penultimate spring rupture is identified.

The springs closest to the lower side of the system (\emph{cf.}
Fig.~\ref{fig:parallell_model}) are subject to compressive stresses
during the major part of the breakdown process. Assuming that the
compressive stresses are sufficiently large, it is possible to
identify the next to last rupture as $m = N - 1$. Carrying out the
limit analysis leads to $\tau = 3$. Hence, the analytic approximation
for $G$ in the limit $\phi \approx 0$ becomes

\begin{equation}
  \label{eq:analytic_low}
  G_{\phi\approx 0} = \frac{1}{4} \phi^3 \sim |\phi|^3.
\end{equation}




\section{Results}
\label{sec:result}

\begin{figure}[htbp]
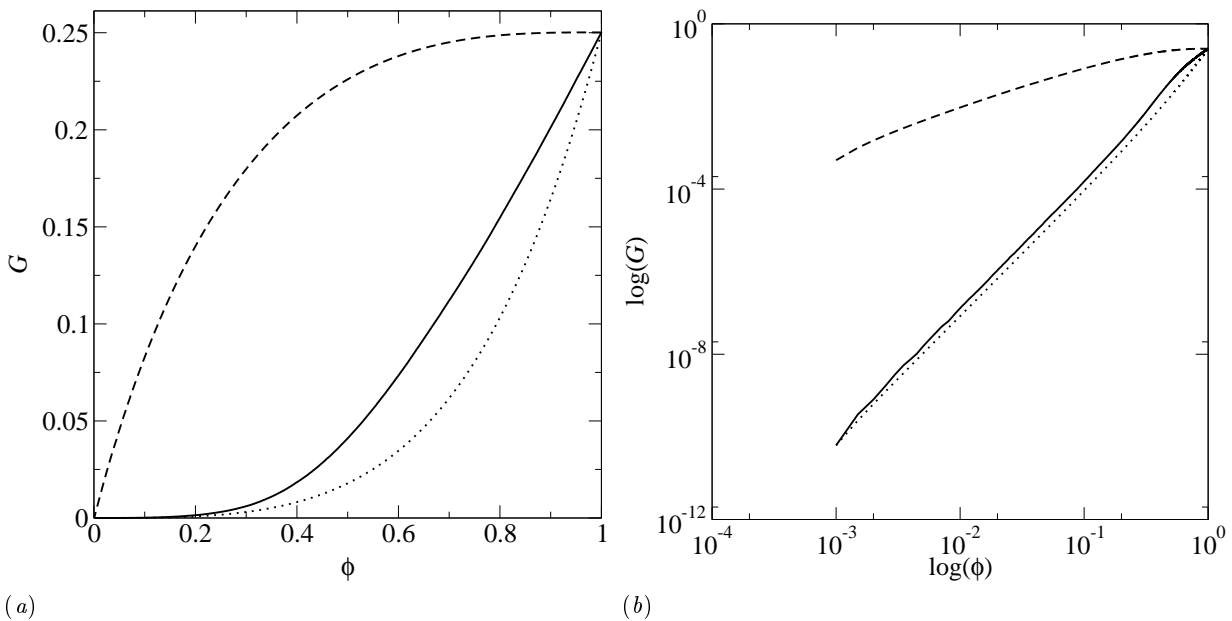

  \centering
  \begin{tabular}[c]{ll}
    \includegraphics[width=0.45\textwidth]{fig/shear_modulus} &
    \includegraphics[width=0.45\textwidth]{fig/shear_modulus_loglog}\\
    (\emph{a}) & (\emph{b})
  \end{tabular}
  \caption{Numerical simulation of $G(\phi)$ averaged over 100 randomly ordered series of
    ruptures of 2001 springs, solid line. The dotted and dashed lines
    depict the lower and upper bounds of $G$, respectively. (\emph{a})
    and (\emph{b}) show linear and logarithmic plots, respectively.}
  \label{fig:shear_modulus}
\end{figure}

The results of our computations of the shear stress, the order
parameter vs. the applied shear force and the burst size distribution
are displayed in a number of figures.
Figures~\ref{fig:shear_modulus}\emph{a-b} show the shear modulus
averaged over 100 rupture sequences of 2001 springs. The sequences
were generated according to the procedure described in
Sec.~\ref{sec:numerical_method}. This curve, henceforth, is called the
average shear modulus curve. These figures also show the lower and and
upper bounds of $G(\phi)$, respectively. The lower bound curve
corresponds to the breaking of the highest strained spring and the
upper curve to the breaking of the least strained spring. From these
figures, it can be deduced that $G(\phi)$ depends linearly on $\phi$
close to $\phi = 1$, while it has a scaling behaviour (power law
dependence) close to $\phi = 0$, \emph{cf.}  Eq.~(\ref{eq:G_0}). Note
that the average and lower bound shear modulus curves coalesce as
$\phi \rightarrow 0$, see Fig.~\ref{fig:shear_modulus}\emph{b}. This
validates the assumption that spring $N-1$ is the second last to break
and consequently that Eq.~(\ref{eq:analytic_low}) is correct.

Using regression, linear and power law expressions are fitted to the
computed data. Here, the intervals $\phi \in [ 0^+; 0.25 ]$ and $\phi
\in [ 0.75; 1.0 ]$ are chosen to determine the power law and linear
law, respectively. Our analysis leads to the following expressions for
$G(\phi)$,

\begin{eqnarray}
  \label{eq:curve_fitting_0}
  G_{\phi\approx 0} & \sim & 0.217 \phi^{3.1297},  \\
  \label{eq:curve_fitting_1}
  G_{\phi\approx 1} & \sim & -0.22259 + 0.47154 \phi.  
\end{eqnarray}

\begin{figure}[htbp]
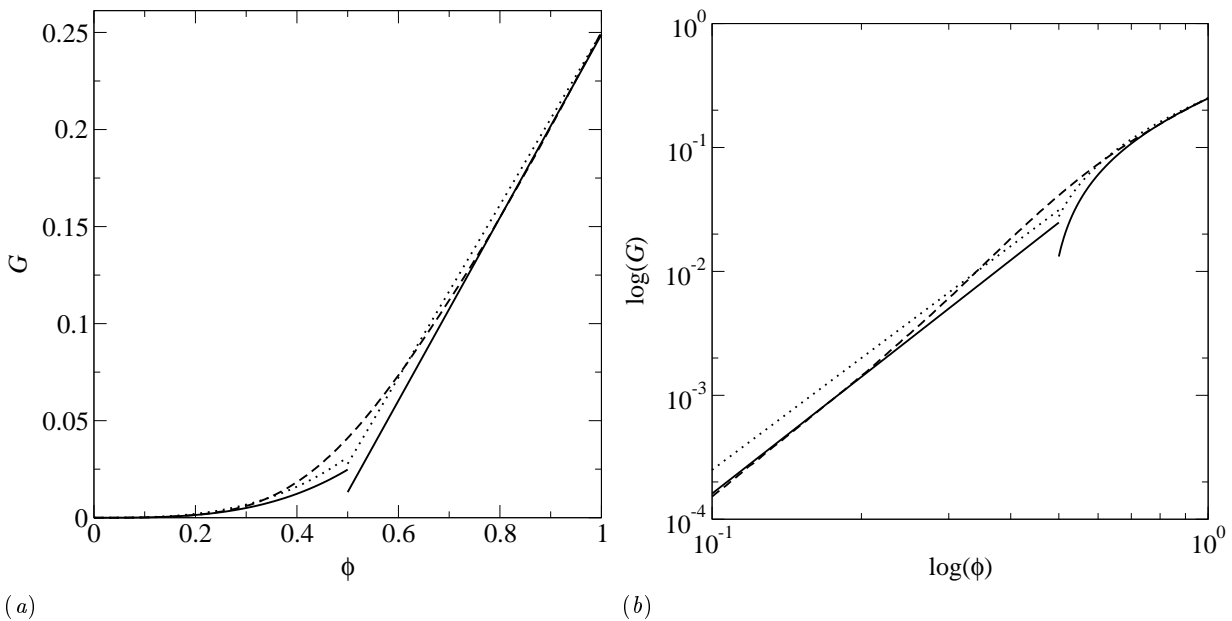

  \centering
  \begin{tabular}[c]{ll}
    \includegraphics[width=0.45\textwidth]{fig/approximation} &
    \includegraphics[width=0.45\textwidth]{fig/approximation_loglog}\\
    (\emph{a}) & (\emph{b})
  \end{tabular}
  \caption{Approximate computations of the shear modulus $G(\phi)$. The solid and dotted
    lines display numerical and analytical approximations,
    respectively. The broken line shows the curve found from averaging
    over several sequences of spring ruptures, see Fig.
    \ref{fig:shear_modulus}.}
  \label{fig:approximation} 
\end{figure}

Figures~\ref{fig:approximation}\emph{a}-\emph{b} show the analytical
(Eqs.~(\ref{eq:analytic_low}) and (\ref{eq:analytic_high})) and
numerical (Eqs.~(\ref{eq:curve_fitting_0})-(\ref{eq:curve_fitting_1}))
scaling and linear laws, respectively, superimposed on the average
shear modulus curve. As can be expected, the approximated curves
deviate for $\phi$ in the mid-range ($\phi \sim 0.5$).

\begin{figure}[htbp]
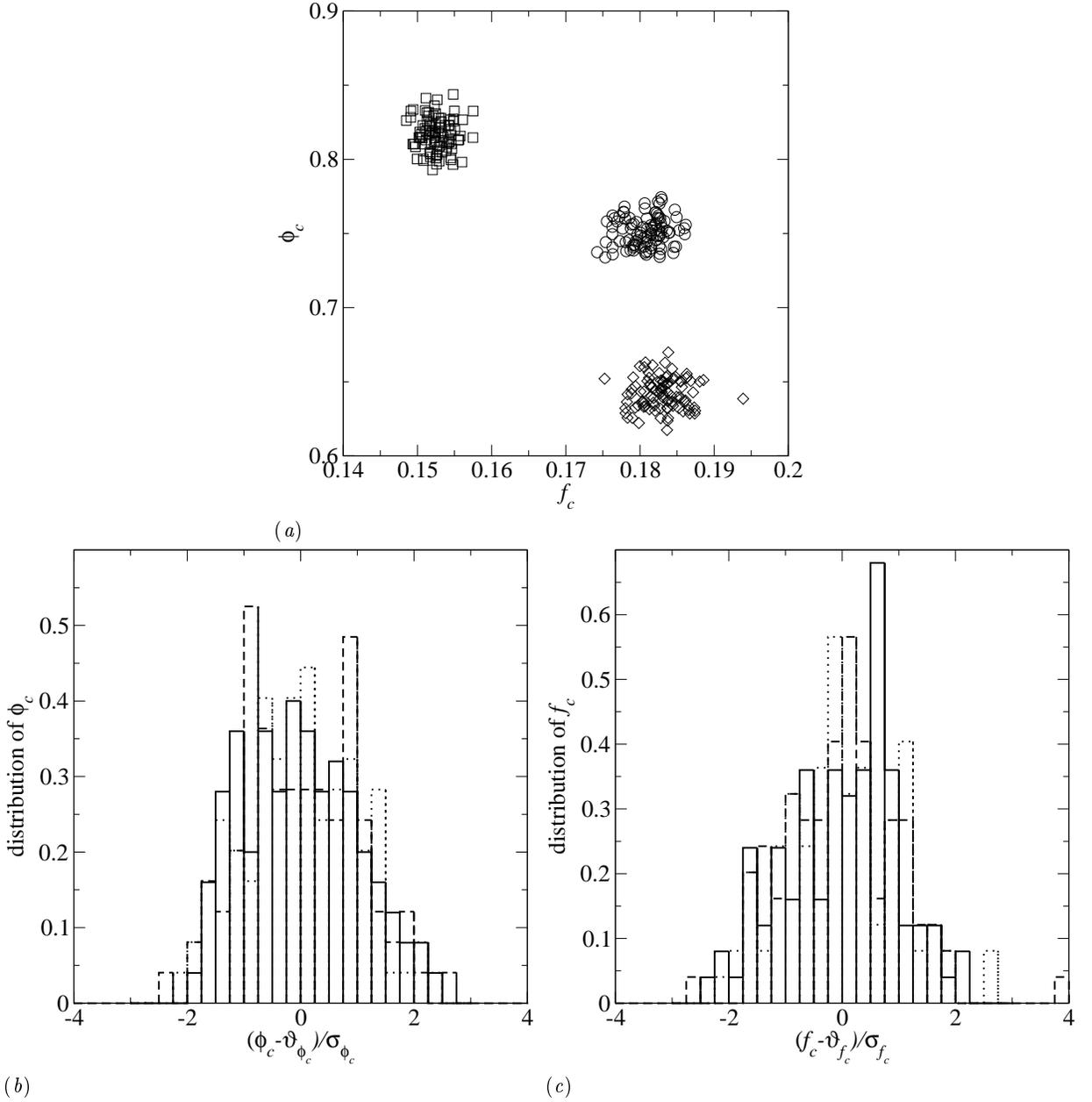

  \centering
  \begin{tabular}[c]{l}
    \includegraphics[width=0.45\textwidth]{fig/phic_v_fc} \\
    (\emph{a})
  \end{tabular}
  \begin{tabular}[c]{ll}
    \includegraphics[width=0.45\textwidth]{fig/phic_distribution} &
    \includegraphics[width=0.45\textwidth]{fig/fc_distribution} \\
    (\emph{b}) & (\emph{c})
  \end{tabular}
  \caption{(\emph{a}) The critical scaling variables $\phi_c$ versus $f_c$ for
    100 configurations, computed by the quasi-static method, of
    uniformly $r_i \in [ 0; 2 ]$, Eq.~(\ref{eq:uniform}), circles,
    Weibull $(\nu,\lambda)$=(2,1), Eq.~(\ref{eq:weibull}), squares and
    normal $(\vartheta,\sigma)$=(1,1), Eq.~(\ref{eq:gauss}), diamonds,
    distributed rupture thresholds of $N = 10001$ springs. (\emph{b})
    and (\emph{c}) display normalised histogram distributions of
    $\phi_c$ and $f_c$, respectively. Solid, dotted and dashed lines
    refer to the uniform, Weibull and normal distribution,
    respectively. The abscissae have been rescaled with the mean
    value, $\vartheta$, and the standard deviation, $\sigma$, to
    enable comparison between results for the three rupture threshold
    distributions.}
  \label{fig:phic_v_fc}
\end{figure}

A system size of $N = 10001$ springs and a load increment of $\Delta f
= 0.01/N$ are used in the numerical simulations (the quasi-static
method discussed in the foregoing section) presented here. To gather
enough data for statistical predictions, each simulation is repeated
for $M = 100$ different configurations of rupture thresholds $r$.
Values from the last load step before global breakdown are taken as
critical values. Figure \ref{fig:phic_v_fc}\emph{a} displays the
simulated critical values of the order parameter, $\phi_c$ and the
shear force $f_c$, for the uniform, Weibull and normal distribution
functions defined in Eqs.~(\ref{eq:uniform})-(\ref{eq:gauss}). The
three distributions display similar results with respect to scatter in
$f_c$ and $\phi_c$, see Figs.~\ref{fig:phic_v_fc}\emph{b}-\emph{c}.
Using these results (average values of $G(\phi_c)$, $\phi_c$
and $f_c$), we have determined $\phi_0$ in Eq. (\ref{eq:G_critical})
for the uniform failure distribution given by Eq. (\ref{eq:uniform}).
We found $\phi_0=1.042$.

\begin{figure}[htbp]
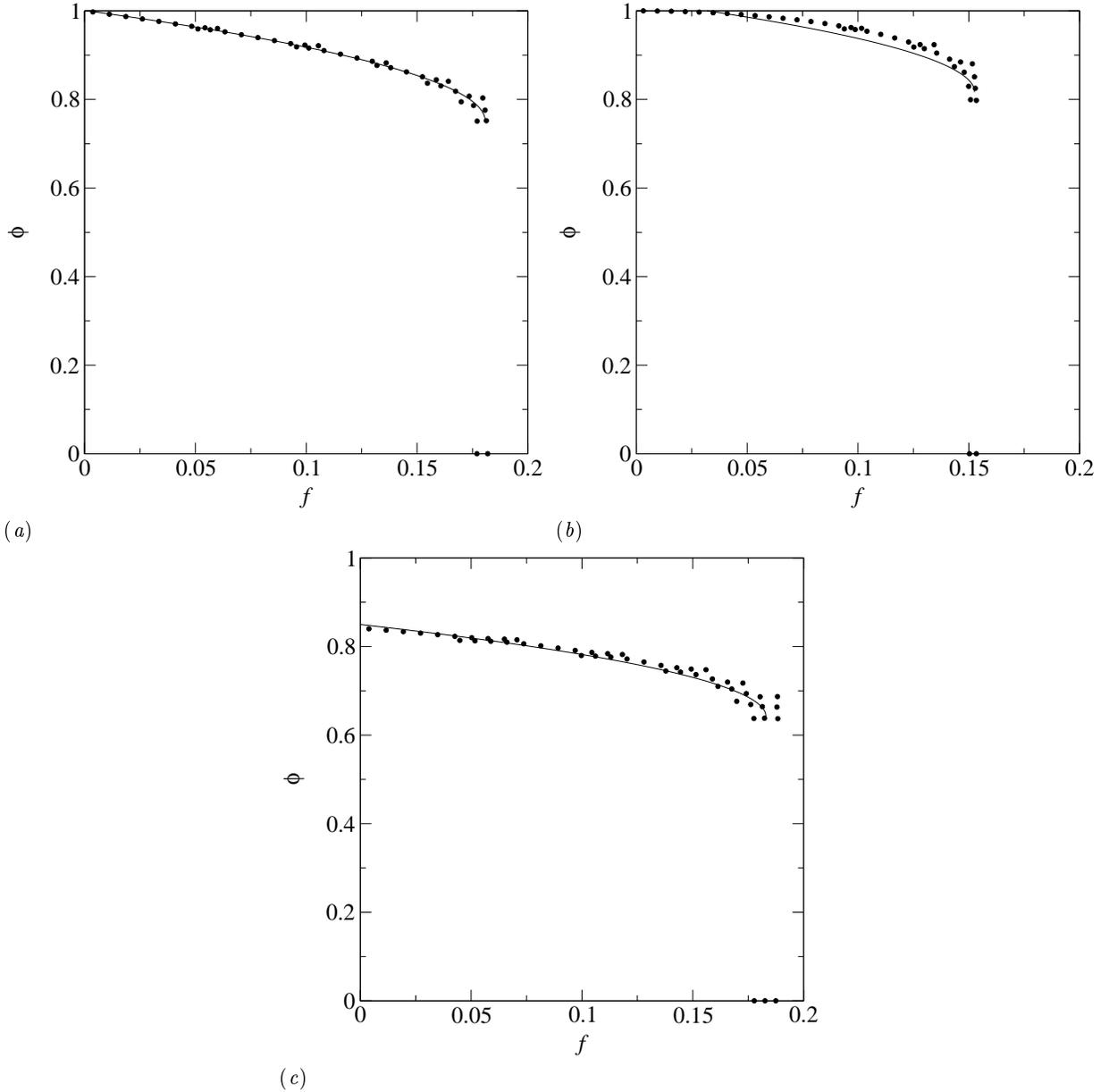

  \centering
  \begin{tabular}[c]{ll}
    \includegraphics[width=0.45\textwidth]{fig/phi_v_f-uni-N10001-M100} &
    \includegraphics[width=0.45\textwidth]{fig/phi_v_f-wei-N10001-M100} \\
    (\emph{a}) & (\emph{b})
  \end{tabular}
  \begin{tabular}[c]{l}
    \includegraphics[width=0.45\textwidth]{fig/phi_v_f-gau-N10001-M100} \\
    (\emph{c})
  \end{tabular}
  \caption{The damage order parameter $\phi$ versus the shear force
    $f$ averaged over 100 configurations, computed by the quasi-static
    method of (\emph{a}) uniform $r_i \in [ 0; 2 ]$, (\emph{b})
    Weibull $(\nu,\lambda)$=(2,1), and (\emph{c}) normal
    ($\vartheta,\sigma$)=(1,1), distributions of rupture thresholds
    for $N = 10001$ springs, respectively. The lines represent the
    mean-field scaling behaviour.}
  \label{fig:phi_v_f}
\end{figure}

Next, we focus on the relationship between the scaling variables
$\phi$ and $f$.  Figures~\ref{fig:phi_v_f}\emph{a}-\emph{c} show the
computed order parameter, $\phi$, plotted against the applied load,
$f$, for the three distributions using the quasi-static method. The
data presented in the figures are averaged over 100 configurations of
rupture threshold. The displayed $(\phi,f)$-plane is divided into
rectangles. The arithmetic mean of the values confined within each
rectangle is used to represent the data in the figures. The number of
points within each rectangle is utilised as weighting factors to fit a
scaling law expression Eq.~(\ref{eq:par_scale}) to the simulated data.
The scaling form of the order parameter predicted by the mean-field
theory Eq.~(\ref{eq:par_scale}) is also depicted in
Figs.~\ref{fig:phi_v_f}\emph{a}-\emph{c}. The simulated data support
the validity of the mean-field analysis.

\begin{figure}[htbp]
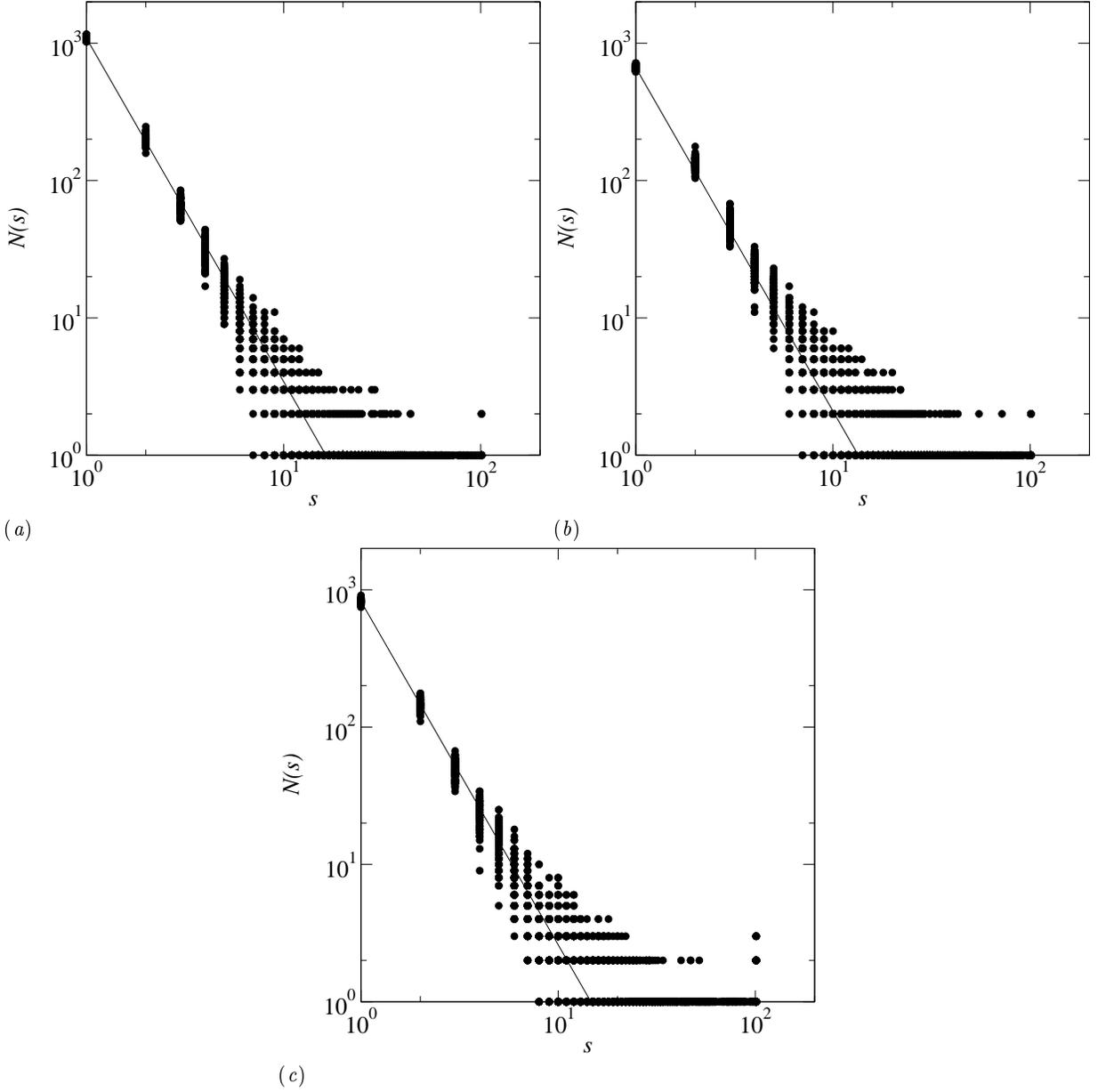

  \centering
  \begin{tabular}[c]{ll}
    \includegraphics[width=0.45\textwidth]{fig/size_distribution-uni-N10001-M100} &
    \includegraphics[width=0.45\textwidth]{fig/size_distribution-wei-N10001-M100} \\
    (\emph{a}) & (\emph{b})
  \end{tabular}
  \begin{tabular}[c]{l}
    \includegraphics[width=0.45\textwidth]{fig/size_distribution-gau-N10001-M100} \\
    (\emph{c})
  \end{tabular}
  \caption{Burst size distributions for 100 configurations using (\emph{a})
    uniform $r_i \in [ 0; 2 ]$, (\emph{b}) Weibull
    $(\nu,\lambda)$=(2,1) and (\emph{c}) normal
    ($\vartheta,\sigma$)=(1,1), probability distributions of rupture
    thresholds for $N = 10001$ springs, respectively. The straight
    lines have a slope of $-2.5$, see Eq.~(\ref{eq:burst_critical}).}
  \label{fig:size_distribution}
\end{figure}

Finally, the important concept of burst distribution is explored.  For
a system undergoing wear, the size of the avalanches can be used to
estimate the size of the wear fragments \cite{Rabinowicz_1995}. The
size of an avalanche or burst, $s$, is defined as the number of
springs that break simultaneously for a constant load
\cite{Batrouni_et_al_2001}. The burst size distribution is generated
by counting the number of bursts of size, $N(s)$, occurring as the
system breaks down.
Figures~\ref{fig:size_distribution}\emph{a}-\emph{c} show the burst
distributions for the three chosen rupture threshold distributions for
the damage model considered here. The three figures display similar
behaviour. However, our simulations indicate that the Weibull and
normal distribution give a higher amount of large bursts ($s>10$). The
figures also include a line with a slope $-5/2$ (log-log scales),
which is the theoretical result found for a fibre bundle model with
global load sharing \cite{Hemmer_Hansen_1992,Hansen_Hemmer_1994}. The
results indicate that this scaling law has relevance for the system
studied here.

\section{Discussion}
\label{sec:discus}

The results presented in the preceding section show that the average
shear modulus is linearly proportional to the damage in the beginning
of the breakdown process and has power law characteristics at end of
the process, see Fig.~\ref{fig:shear_modulus}.  Consequently, the
shear modulus can be approximated with simple relations, which do not
require knowledge of the path to failure, \emph{i.e.} the order in
which the springs break.  The critical force and critical order
parameter (damage) can be estimated analytically using mean-field
theory once the shear modulus has been determined, provided that the
distribution of the disorder of the material is known. Here, it is
assumed that the disorder is quenched in the system and does not
change during the breakdown process \cite{Hermann_Roux_1990}.

In the preceding sections, we showed that for $\phi \approx 0$, the
shear modulus $G$ scales as $G \sim \phi^\tau$, with $\tau=3$, which
we obtained by analytical calculation, while numerical simulations led
to $\tau \approx 3.13$ in the interval $\phi \in [ 0^+; 0.25 ]$. These
results may be compared with properties of other condensed matter
systems, \emph{e.g.} the macroscopic conductance in a resistor network
or the elastic modulus of gels in polymer-system
\cite{de_Gennes_1976}. For lattice dimensions $d \ge 6$, which
corresponds to a mean-field approximation, de Gennes found that $\tau
=3$.  This result also conforms with a recent calculation by Xing
\emph{et al.} of the scaling behaviour of shear modulus in
polymer-system \cite{Xing_et_al_2004}. Their renormalisation group
analysis shows that $\tau=3-\frac{5}{21}\epsilon$, with
$\epsilon=6-d$, which is identical to the conductivity exponent of the
random resistor network calculated by Harris and Lubensky
\cite{Harris_Lubensky_1987}.

The constitutive relation for damage, $\phi$ vs. $f$, for all the
three statistical distributions (Figs.
\ref{fig:phi_v_f}\emph{a}-\emph{c}) indicates that $\phi$ decreases
slowly with the increase in $f$, then a sharp (discontinuous) drop in
$\phi$ occurs, when the shear force reaches a threshold value $f_c$.
This sudden transition in the order parameter and the fact that the
scaling form given by Eq.~(\ref{eq:par_scale}) describes the approach
to $f_c$ so well ($f <f_c$), indicates that material breakdown is a
first-order phase transition, \emph{e.g.}  similar to a spinodal
decomposition, see Appendix.

The burst distributions for the system using the three statistical
distributions are displayed in 
Figs.~\ref{fig:size_distribution}\emph{a}-\emph{c}.  Here the size
distribution of avalanches is integrated over the load.  For a fibre
bundle model (FBM) with global load sharing (the load is evenly
distributed over the remaining fibres) the expected number of
avalanches $\mathcal{N}(s)$, where $s$ is the size of the avalanche,
scales as $\mathcal{N} \sim s^{-\xi}$, with $\xi = 5/2$
\cite{Hemmer_Hansen_1992,Hansen_Hemmer_1994}.  Although this scaling
law holds for a large class of threshold distributions, this is not
always the case \cite{Kloster_et_al_1997}.  For small avalanches, $s <
10$, a regression analysis, using the numerical data shown in the
figures, gives an exponent close to the FBM value, $2.5 \pm 0.2$. It
may be worth to compare our results with that for the random-field
Ising model (RFIM) predictions, which intends to describe the
behaviour of many magnetic materials subjected to external magnetic
field $H$. For this system, the magnetisation changes through
nucleation and motion of domain walls. The motion is discrete,
\emph{i.e.}  there will be magnetisation bursts corresponding to
reorganisation (or avalanche) of a domain of spins which span several
decades of size \cite{Perkovic_et_al_1995}. There will be plenty of
small avalanches of spins and fewer and fewer avalanches of larger and
larger sizes \cite{Sethna_et_al_2004}. The power law dependence for
the probability of having an avalanche of a given size has been
determined \cite{Perkovic_et_al_1995}. It is shown that the size
distribution of all avalanches that occur in one branch of hysteresis
loop (for $H$ from $-\infty$ to $+\infty$) scales in the form:
$\mathcal{N}_{int} \sim s^{-\theta}\mathcal{F}(s,r)$, where
$\mathcal{F}$ is a scaling function and $r$ is a ``disorder''
parameter \cite{Perkovic_et_al_1999}. The mean-field computation of
Perkovi\'{c} \emph{et al.} \cite{Perkovic_et_al_1999} gave $\theta =
\frac{9}{4}$, which is close to the corresponding value predicted by
FBM.
 
The breakdown of the system that we considered is similar to that of
FBM.  The process of breakdown is envisaged to proceed by the
formation of micro-cracks. The micro-cracks are nucleated according to
a linear distribution function, which is proportional to the
displacement field. The cracks grow up to a critical size and coalesce
to form the final crack. We found no indication that a
single crack takes over and dominates the breakdown process. This
behaviour was also found in a two dimensional mesh subject to a uniform
stress field \cite{Zapperi_et_al_1999}.



\section{Conclusion}
\label{sec:conclude}

We modelled an interface layer by $N$ parallel elastic springs that
connected two rigid blocks.  We loaded the system by a shear force $f$
acting on the top side. The springs were assumed to have equal
stiffness but were ruptured randomly when the load reached a critical
value $f_c$. The state of the material damage for the system, was
characterised by an order parameter, $\phi$, varying between zero
(complete breakdown) and one (intact). The shear modulus for the
system was numerically simulated over the entire range of $\phi$ and
it was found that $G_{\phi\approx 0} \sim \phi^{3.13}$ and
$G_{\phi\approx 1} \sim \phi$. The linear behaviour of $G(\phi)$ close
to $\phi = 1$ was confirmed analytically. Moreover, our analytical
calculation showed that $G_{\phi\approx 0} \sim \phi^{3}$. We
calculated $\phi$ vs. $f$ numerically and determined their values at
the onset of breakdown, \emph{i.e.}  around $(\phi_c, f_c)$, and also
found their variation depending on the failure distribution function
selected. We noted that there was a discontinuous drop in $\phi$
around $f = f_c$.  The mean-field theory predicted this behaviour for
$f \lessapprox f_c$, leading to the conclusion that breakdown behaves
like a first-order phase transition.  Finally, we calculated the burst
size distribution during rupture and found that the system behaves
according to the predictions of the fibre bundle model.
 
The model presented here can be extended to treat a more realistic
representation of wear induced loads by modelling a two-dimensional
lattice comprising a network of blocks connected to each other and to
a rigid indenter by elastic springs.


\begin{acknowledgments}
  The work was supported by the Swedish Foundation for Knowledge and
  Competence Development (KKS) under Award HÖG 212/01.
\end{acknowledgments}

\appendix*
\section{Magnetic system and lattice damage: an analogy}
\label{sec:append}

In magnetic language (Ising model), let us envisage that the lattice
sites confine ``spins'' which may have only values $\sigma_i=+1$ and
$\sigma_i=-1$. These spins interact with the nearest-neighbours on the
lattice with a constant interaction energy $J_{ij}$, see \emph{e.g.}
\cite{Baxter_1982}. In particular, we consider the random field Ising
model (RFIM), where at each site $i$ there is a random field $g_i$ and
there is an external field $H$ acting on the system, the Hamiltonian
is given by \cite{Sethna_et_al_1993}

\begin{equation}
  \label{eq:rfim_hamilton}
  \mathcal{H}_I = -\sum_{<ij>}^N J_{ij}\sigma_i \sigma_j-\sum_{i}^N
  (H+g_i)\sigma_i.
\end{equation}

The system transforms from negative to positive magnetisation as the
external field $H$ is swept upward. Supposing that each spin
$\sigma_i$ always gets aligned with the effective local field and
$J_{ij}=J$, the Hamiltonian (\ref{eq:rfim_hamilton}) is simplified to

\begin{equation}
  \label{eq:eff_field}
  \mathcal{H}_I = -\sum_{i}^N \sigma_i\Big(\sum_{j}^N J\sigma_j+H+g_i\Big).
\end{equation}

In the mean-field theory of Weiss
(\cite{Baxter_1982},\cite{Goldenfeld_1992}), one assumes that every
spin is coupled to all $N-1$ other spins with a coupling constant
$J/(N-1)$. The effective field acting on site $i$ is now expressed as
$h_i=Jz\mathcal{M}/(N-1)+H+g_i$ where $z$ is the lattice coordination
number and $\mathcal{M}$ is the magnetisation, the order parameter,
defined as

\begin{equation}
  \label{eq:order_param}
  \mathcal{M} = \sum_{j}^N\sigma_j=N_+ - N_- ,
\end{equation}
\noindent
where $N_+=(N+\mathcal{M})/2$ and $N_-=(N-\mathcal{M})/2$ are the
number of up-spins and down-spins, respectively. Moreover, spins with
$g_i<-(Jz\mathcal{M}/(N-1)+H)$ will point downward, others will point
upward. Since $g_i$ is a stochastic variable, the self-consistency
\cite{Sethna_et_al_1993} provides

\begin{equation}
  \label{eq:order_param_2}
  \mathfrak{m}(H) = 1 - 2\int_{-\infty}^{-Jz\mathfrak{m}-H} \rho(g)\mathrm{d}g,
\end{equation}
\noindent
where $\mathfrak{m}=\mathcal{M}/N$ is the magnetisation per site
($N>>1$) and $\rho(g)$ is the probability distribution for the random
field $g$. The magnetic susceptibility is calculated by
differentiating Eq. (\ref{eq:order_param_2}) with respect to $H$, viz.

\begin{equation}
  \label{eq:mag_chi}
  \chi = \frac{d\mathfrak{m}}{dH} = \frac{2\rho(x)}{1-2J\rho(x)},
\end{equation}
\noindent
where $x=-J\mathfrak{m}-H$ and we put $z=1$ for convenience. Note that
at $x_c=\rho^{-1}(0.5/J)$ the susceptibility diverges. Taylor
expansion around $x_c$ yields,
\begin{equation}
  \label{eq:mag_chi_exp}
  \chi =  J^{-1}+
  \frac{0.5J^{-2}}{\rho'(x_c)(x-x_c)+\frac{1}{2}\rho''(x_c)(x-x_c)^2+\cdot\cdot\cdot}.
\end{equation}

Near $x_c=-J\mathfrak{m}_c-H_c$, we have
\begin{equation}
  \label{eq:mag_chi_approx}
  \chi \approx \frac{0.5J^{-2}}{\rho'(x_c)(x-x_c)}.
\end{equation}

Re-writing Eq. (\ref{eq:mag_chi_approx}) in terms of $\mathfrak{m}$
and $H$,
\begin{equation}
  \label{eq:mag_chi_de}
  \frac{d\mathfrak{m}}{dH} = -\frac{A}{J(\mathfrak{m}-\mathfrak{m}_c)+(H-H_c)},
\end{equation}
\noindent
with $A=0.5J^{-2}/\rho'(x_c)$. Integrating Eq. (\ref{eq:mag_chi_de}),
we find the equation of the state in transcendental form for $H<H_c$
and, close to $H_c$, viz.
\begin{equation}
  \label{eq:mag_sol}
  \mathfrak{m}-\mathfrak{m}_c = -B\log{[1+ \frac{\mathfrak{m}_c-\mathfrak{m}+H_c-H}{B}]},
\end{equation}
\noindent
with $B=JA$. Again series expansion of Eq. (\ref{eq:mag_sol}) around
$\mathfrak{m}_c$ and $H_c$ results in the dominating contribution,
\begin{equation}
  \label{eq:mag_asy_sol}
  \mathfrak{m}-\mathfrak{m}_c \sim (H_c-H)^{1/2}.
\end{equation}

The susceptibility in this region ($H \approx H_c$) scales as
\begin{equation}
  \label{eq:mag_chi_asy_sol}
  \chi \sim (H_c-H)^{-1/2}.
\end{equation}

The scaling behaviour for the magnetisation, Eq.
(\ref{eq:mag_asy_sol}) and the susceptibility Eq.
(\ref{eq:mag_chi_asy_sol}), close to $H_c$ for $H<H_c$, resemble that
of the spinodal decomposition in the mean-field theory of nucleation
\cite{Klein_Unger_1983}.

For the lattice damage model, upon analogy with RFIM, we write the
following Hamiltonian

\begin{equation}
  \label{eq:damage_hamilton_1}
  \mathcal{H}_D = \sum_{<ij>}^N \epsilon \kappa_i\kappa_j +\sum_{i}^N \kappa_i(\mu-p_i),
\end{equation}
\noindent
where $\epsilon$ is the inter-spring interaction parameter, $\mu$ is
the effective chemical potential and $p_i$ is a random field variable.
To make contact between the lattice damage and magnetic system, we
write: $\sigma_i = 2\kappa_i-1$. The lattice damage Hamiltonian, Eq.
(\ref{eq:damage_hamilton_1}), expressed in terms of spin variable
$\sigma_i$, becomes

\begin{equation}
  \label{eq:damage_hamilton_2}
  \mathcal{H}_D = \frac{1}{4}\Big[ \epsilon \sum_{<ij>}^N
 \sigma_i\sigma_j
  +2(\epsilon + \mu)\sum_{i=1}^N\sigma_i - 2\sum_{i=1}^Np_i\sigma_i\Big],
\end{equation}
\noindent
where we have left out the constant spin-independent terms.  Comparing
Eqs.  (\ref{eq:damage_hamilton_2}) and (\ref{eq:rfim_hamilton}), we
can relate the variables in the magnetic model and the lattice damage
model as follows
 
\begin{equation}
  \label{eq:energy_analog_1}
  J=-\epsilon/4; \qquad H =-(\epsilon+\mu)/2; \qquad g_i=p_i/2.
\end{equation}

Let us first rewrite Eq. (\ref{eq:damage_hamilton_1}) in the form
\begin{equation}
  \label{eq:damage_hamilton_mf1}
  \mathcal{H}_D = \sum_{i}^N \kappa_i\Big(\epsilon \sum_{j}^N\kappa_j +\mu-p_i\Big),
\end{equation}
\noindent
then introduce the order parameter $\Phi$ for the lattice damage model

\begin{equation}
  \label{eq:damage_order2}
  \Phi = \sum_{i}^N \kappa_i= N-q.
\end{equation}

Here, ($N-q$) and $q$ are the number of intact and broken springs,
respectively. In the mean-field theory Eq.
(\ref{eq:damage_hamilton_mf1}) is replaced by

\begin{equation}
  \label{eq:damage_hamilton_mf2}
  \mathcal{H}_D = \sum_{i}^N \kappa_i\Big(\frac{z\epsilon\Phi}{N-1} +\mu-p_i\Big),
\end{equation}

Fashioned after Zapperi \textit{et al.}
\cite{Zapperi_et_al_1997,Zapperi_et_al_1999}, we rewrite Eq.
(\ref{eq:energy_force}) in the form

\begin{equation}
  \label{eq:mean_field_3}
  \mathcal{E} =  \frac{1}{2} \sum_{i=1}^N \kappa_i 
  \left( \frac{Nf^2}{\Phi G(\Phi)} - r_i ^2 \right), 
\end{equation}
\noindent
where $G(\Phi)$ is re-expressed as a function of the order parameter
$\Phi$. Now comparing Eqs. (\ref{eq:damage_hamilton_mf2}) and
(\ref{eq:mean_field_3}), we identify

\begin{equation}
  \label{eq:energy_analog_2}
  z\epsilon\phi+\mu= \frac{f^2}{2\phi G(\phi)}; 
\qquad r_i^2/2 = p_i,
\end{equation}
\noindent
where we placed $\phi=\Phi/N$.

The relation between $\phi$ and the magnetic order parameter is
$\mathfrak{m}=2\phi-1$. Using this relation and Eq.
(\ref{eq:energy_analog_1}), we can write Eq.  (\ref{eq:order_param_2})
as

\begin{equation}
  \label{eq:order_param_3}
  \phi = 1 -\int_{-\infty}^{\epsilon\phi+\mu/2} \rho(g)\mathrm{d}g
  = 1- \frac{1}{2}\int_{0}^{2\epsilon\phi+\mu} \rho(p)\mathrm{d}p,
\end{equation}
\noindent
where we placed $z=2$ and considered only $p_i>0$. By invoking Eq.
(\ref{eq:energy_analog_2}), we find

\begin{equation}
  \label{eq:order_param_5}
  \phi = 1-\frac{1}{2}\int_{0}^{\frac{f}{\sqrt{\phi G(\phi)}}} \rho(r)r\mathrm{d}r.
\end{equation}

Similar calculations as made on the magnetic order parameter
$\mathfrak{m}$ yield the scaling relations for the lattice damage
order parameter $\phi$ and the susceptibility $\chi$, \emph{cf.} Eqs.
(\ref{eq:mag_asy_sol}) and (\ref{eq:mag_chi_asy_sol}); \emph{i.e.}
close to $\phi_c$ for $f<f_c$, we have \cite{Zapperi_et_al_1999},
\begin{eqnarray}
  \phi-\phi_c &\sim& (f_c-f)^{1/2}, \label{eq:para_lattice_critical}\\
  \chi = \frac{d\phi}{df} &\sim& (f_c-f)^{-1/2}, \label{eq:chi_lattice_critical}
\end{eqnarray}
\noindent
where $f_c$ is the critical force for the lattice breakdown. Both
relations (\ref{eq:para_lattice_critical}) and
(\ref{eq:chi_lattice_critical}) are the mean-field theory
predictions.




\end{document}